\documentclass[acmlarge,authorversion,nonacm]{acmart}
\AtBeginDocument{%
  \providecommand\BibTeX{{%
    \normalfont B\kern-0.5em{\scshape i\kern-0.25em b}\kern-0.8em\TeX}}}

\usepackage{multirow}
\usepackage{threeparttable}
\makeatletter
\def\@copyrightspace{\relax}
\makeatother

\begin{document}

%\title{Speed and Accuracy of Skill Tagging Through AI-Human Collaboration}
\title{Human-AI Collaboration Increases Skill Tagging Speed but Degrades Accuracy}


% \author{Anonymous}
\author{Cheng Ren}
\authornotemark[1]
\email{cren@albany.edu}
\affiliation{%
  \institution{University at Albany, State University of New York}
  \country{United States}
}
\author{Zachary Pardos}
\authornotemark[2]
\email{pardos@berkeley.edu}
\affiliation{%
  \institution{University of California, Berkeley}
  \country{United States}
}
\author{Zhi Li}
\authornotemark[2]
\email{zhili@berkeley.edu}
\affiliation{%
  \institution{University of California, Berkeley}
  \country{United States}
}

% \affiliation{%

% \affiliation{%

% \affiliation{%

% \affiliation{%

% \affiliation{%

% \affiliation{%

% \affiliation{%

\renewcommand{\shortauthors}{Ren, et al.}

\begin{abstract}
AI approaches are progressing besting humans at game-related tasks (e.g. chess). The next stage is expected to be Human-AI collaboration; however, the research on this subject has been mixed and is in need of additional data points. We add to this nascent literature by studying Human-AI collaboration on a common administrative educational task. Education is a special domain in its relation to AI and has been slow to adopt AI approaches in practice, concerned with the educational enterprise losing its humanistic touch and because standard of quality is demanded because of the impact on a person's career and developmental trajectory. In this study (N = 22), we design an experiment to explore the effect of Human-AI collaboration on the task of tagging educational content with skills from the US common core taxonomy. Our results show that the experiment group (with AI recommendations) saved around 50\% time (p $<$$<$ 0.01) in the execution of their tagging task but at the sacrifice of 7.7\% recall (p = 0.267) and 35\% accuracy (p= 0.1170) compared with the non-AI involved control group, placing the AI+human group in between the AI alone (lowest performance) and the human alone (highest performance). We further analyze log data from this AI collaboration experiment to explore under what circumstances humans still exercised their discernment when receiving recommendations. Finally, we outline how this study can assist in implementing AI tools, like ChatGPT, in education.
\end{abstract}






\maketitle

\section{Introduction}
AI has made significant advancements and achieved impressive performance in a variety of tasks across multiple fields. The rapid progress of AI in numerous benchmarks leads to the expectation that it will incraesingly be able to perform tasks in real-world settings. For instance, a meta-analysis conducted by Liu et al. \cite{liu2019comparison} compared the diagnostic performance of deep learning algorithms to healthcare professionals in detecting diseases from medical imaging, and found that the two were similar. Similarly, Chen et al. \cite{chen2020} systematically reviewed the use of AI in education and found that it has been applied in education administration, instruction, and learning, and has had a somewhat positive impact in these areas. While AI has demonstrated impressive performance in these tasks, it is not currently capable of replacing humans in the decision-making process. Instead, its role is to support humans in decision-making.

The results of studies on Human-AI collaboration in real-world tasks have been mixed. Weisz et al. \cite{weisz2022better} reviewed a set of such studies and found that only two out of ten experiments demonstrated a benefit in both time savings and outcome quality with the introduction of AI \cite{desmond2021increasing,ashktorab2021ai}. Comparing these two studies, which involve categorizing a customer support question,  with the skill tagging task, the skill tagging task is more difficult because of the latent nature of cognitive processes. The process of skill tagging can be likened to linking questions with labels, while also incorporating a deeper comprehension of the procedures necessary to achieve objectives or fulfill tasks. It is uncertain whether the current results in the general computer-human interaction field can be generalized to education. In the domain of education, where there is a need for more efficient processes to embrace AI in a humanistic endeavor, it is important to collect empirical support and study the effect of AI+human interaction on educational tasks. 

Education, as an industry, is motivated to adopt AI for several reasons, such as improving efficiency in administrative tasks, personalizing to students' needs, and enhancing the quality of teaching. However, educational institutions, particularly public schools, are often constrained by limited budgets due to their nonprofit status. According to an estimation by The Century Foundation, K-12 public schools in the US are underfunded by approximately \$150 billion annually \cite{cfundation2020}. As a result, educational institutions are keen to consider more efficient approaches to administration. However, AI adoption in education also raises concerns. One major concern is fairness and discrimination, which are critical issues in AI and are even more significant in education where equity is often a paramount concern \cite{akgun2021artificial,jiang2021towards}. Education is not merely a process of imparting knowledge but also a humanistic pursuit, deeply entwined with the relationship between teachers and students. Even functions that appear predominantly administrative, such as skill tagging, bear considerable weight on students, thereby leading to a sense of responsibility to retain human supervision and maintain a humanistic touch. Additionally, due to the profound implications that education has on determining an individual's future pathway, there is an imperative to keep high-quality standards for the operational performance of AI in educational settings.

Our study aims to contribute a data point on the efficiency and outcome quality of AI+human interaction in skill tagging tasks task. The paper will review related works on AI+human interactions in general as well as in AI in education. It will then introduce the methods used in the study, including data sources, model design, and experiment design. The results of the experiment and a discussion section will follow. Finally, the paper will discuss limitations and future works.

\section{Related Works}
Organizations and individuals often spend a significant amount of time aligning their own standards with those of others. Taxonomy alignment typically requires the manual work of subject matter experts. To make this process more efficient, researchers have turned to AI techniques. For example, Choi et al. \cite{choi2016model} used phrase graphs to calculate the similarity between skills, while Yilmazel et al. \cite{yilmazel2007text} employed rule-based techniques and machine learning to extract and classify features from skill descriptions. Koedinger et al. \cite{rivers2016learning} used learning curve analysis to build models for knowledge components in teaching Python programming. These models are able to map questions to such skills. Li et al.\cite{li2021learning} used problem text, response sequences, and modern neural approaches from computational linguistics for taxonomy mapping. Recently, Shen et al.\cite{shen2021classifying} applied a variant of BERT called "task adaptive BERT" on multi-sources like problem descriptions, skill descriptions for skill classification.

In the educational field, it is essential to note that humans are the final decision makers, even when AI recommendations are provided. Weisz et al. \cite{weisz2022better} concluded that AI+human interactions have mixed outcomes across different domains and even within domains. For example, within the field of education, Cognitive Tutors, Teacher+AI, support can assist students in achieving proficiency comparable to that attained through conventional, teacher-only instruction, accomplishing this in one-third less time \cite{anderson1995cognitive}. Holstein et al. \cite{holstein2018student} observed that AI-enhanced classrooms equipped with real-time analytics about student learning can help narrow the gap in students' learning outcomes. Recently, Weitekamp et al. \cite{weitekamp2020interaction} conducted an experiment using Machine Learning to aid Intelligent Tutoring System (ITS) authoring (Human-AI interaction) in comparison to human authoring utilizing Cognitive Tutor Authoring Tools (CTAT, Human only). The AI-assisted tool demonstrated a 75\% time-saving  in completeness compared to CTAT. Recently, Demszky and Liu \cite{demszky2023m} developed an AI system to provide feedback to instructors on dialogic teaching methods, enhancing mentors'  uptake of student contributions by 10\% and improving the learning experience for students.

In other fields, Lai et al. \cite{lai2019human} only improved the quality of deception detection tasks, while Weber et al. \cite{weber2020draw} found that AI+human did not produce higher quality in image restoration tasks and Clark et al. \cite{clark2018creative} noted that AI+human reduced the quality of creating marketing slogans.

To explore the mechanisms behind these results, scholars have examined interactions from various perspectives. Some studies have observed that individuals might place inappropriate trust in AI due to either over- or under-reliance on it \cite{zhang2020effect,siau2018building}. Dietvorst et al. \cite{dietvorst2015algorithm} discovered that when algorithmic forecasters commit errors identical to those made by human forecasters, people lose confidence in the algorithmic forecasters more swiftly. Additionally, an individual’s trust may be influenced by both the declared and perceived accuracy of the system \cite{yin2019understanding}. When tasks become exceptionally complex, individuals might overestimate the capabilities of AI and elevate their trust in it \cite{chong2022human}. Although investigations in this domain have been conducted from diverse viewpoints, the research is frequently general, and specific educational research—especially research on skill tagging with AI—remains relatively understudied.

Neural networks have been very effective at learning semantics from a variety of data types, such as natural languages and images. These fields have advanced modeling techniques in their ability to learn semantic signals from these data sources. Scholars are using new techniques in AI to help humans accomplish partial or even entire workloads in various real-world scenarios. For example, natural language processing (NLP) techniques have been applied in education on auto grading of short open-end questions, essays, and improving course recommendations just to name a few \cite{dong2017attention,ke2019automated,pardos2019connectionist}. 

\section{Methods}
In this study, we aim to compare the performance of two groups of skill taggers: one group that receives AI assistance (experimental group) and one group that does not (control group). We randomly picked 30 problems from grade 6 on the digital learning platform CK12\footnote{\url{https://www.ck12.org/student/}} to use in our experiment. All problems have text and some have images. We used a pre-trained model from \cite{li2023ai} to generate the Common Core skill tags, a common standard used by many US states, which will be used as AI recommendations.

We will first discuss the AI assistance model that we use, including its data source, training, testing and selecting strategies. Then, we will detail the design of our experiment and explain how we will analyze the results.

\subsection{Machine Learning Model}

We use a pre-trained model from \cite{li2023ai} to generate Common Core skill recommendations for the selected 30 problems. The model is trained on Khan Academy(N=21,475)\footnote{\url{https://www.khanacademy.org/}} problems to Common Core skills(N=385) mapping. The model takes the problem text and associated image as input, generates a text vector via sentenceBERT \cite{reimers2019sentence} and an image encoder via EfficientNet \cite{tan2019efficientnet}, fuses them into a single vector via Compact Bilinear Pooling \cite{gao2016compact}, and maps the vector to the associated skills using either a classification model or a similarity matching model. The classification model is a neural network classifier that is trained to predict skill labels using the problem vector, while the similarity matching model(Translation model) encodes the skills using sentenceBERT and ensures the problem vector and the associated skill vector have a high cosine similarity.

Since pre-trained models\cite{li2023ai} provided many versions to choose from, we also tested those models on similar problems from CK12 to determine the best model to use (i.e., classification model or similarity matching model, and whether to include images in the input). Specifically, we collected 18,728 problems from other grades in CK12 to conduct an offline experiment. We enumerated all 385 Common Core skills, and their skill descriptions, associating to this CK12 content found on their website and consider these the ground truth labels, which were aligned by experts from CK12. We used all grades on CK12 as the test set except grade 6, which is the grade we used for the online study, to avoid overfitting to the problems. The evaluation metric was recall@3, as we will provide 3 recommendations for each question in the online study. The results of the offline experiments, shown in Table \ref{tab:offline}, indicate that the similarity matching model with both text and image input had the highest average recall@3 at 0.496. Therefore, we will use this model to generate recommendations for the online study.

\begin{table}
\vspace{-2mm}
\caption{CK12 problems to Common Core skills mapping recall@3 results by the top level of Common Core (excluding Grade 6). The classification and similarity matching models( are both pre-trained on Khan Academy problems to Common Core skills mapping. The top levels of Common Core skills are Grade 7, Grade 8, HSA (high school algebra), HSF (high school functions), HSG(high school geometry), HSN (high school number and quantity) and HSS (high school statistics). The boldface indicates the model with the best performance}
\label{tab:offline}       % Give a unique label
\begin{tabular}{cccccccccc}
\hline\noalign{\smallskip}
Model & Mode(s) & Grade 7 & Grade 8 & HSA & HSF & HSG & HSN & HSS & Average\\
\noalign{\smallskip}\hline\noalign{\smallskip}
\multirow{3}{*}{{\centering Classification}} & Text  & 0.418 & 0.502 & 0.471 & 0.522 & 0.491 & 0.368 & 0.404 & 0.454 \\
& Image & 0.226 & 0.332 & 0.292 & 0.373 & 0.202 & 0.234 & 0.246 & 0.272 \\
& Text + Image & 0.421 & 0.554 & 0.473 & 0.524 & 0.503 & 0.403 & 0.410 & 0.469\\
\noalign{\smallskip}\hline\noalign{\smallskip}
\multirow{3}{*}{{\centering Translation}} & Text  & 0.478 & 0.540 & 0.547 & 0.487 & 0.517 & 0.436 & 0.427 & 0.490 \\
& Image & 0.204 & 0.269 & 0.295 & 0.316 & 0.389 & 0.137 & 0.275 & 0.269 \\
& Text + Image & 0.482 & 0.547 & 0.523 & 0.516 & 0.543 & 0.426 & 0.434 & \textbf{0.496}\\
\noalign{\smallskip}\hline
\end{tabular}
%\vspace{-6mm}
\end{table}

\subsection{Online Study}

\subsubsection{Experiment Design}
To compare the performance of students with and without AI assistance, we collected responses to a survey using the Qualtrics platform. The survey consisted of 30 math problems, and for each problem, students were asked to select the skills that were demonstrated in the problem. There were five broad categories of skills to choose from: Ratios \& Proportional Relationships, The Number System, Expressions \& Equations, Geometry, and Statistics \& Probability. Once a broad category was chosen, students could select the specific sub-categories of skills within that category. Students were allowed to select multiple skills for each problem, up to a maximum of three. Meanwhile, they could not move to the next question if they only choose the big category but did not choose any skills under it.

\begin{figure}
    %\vspace{-5mm}
    \centering
    \includegraphics[width=0.5\linewidth]{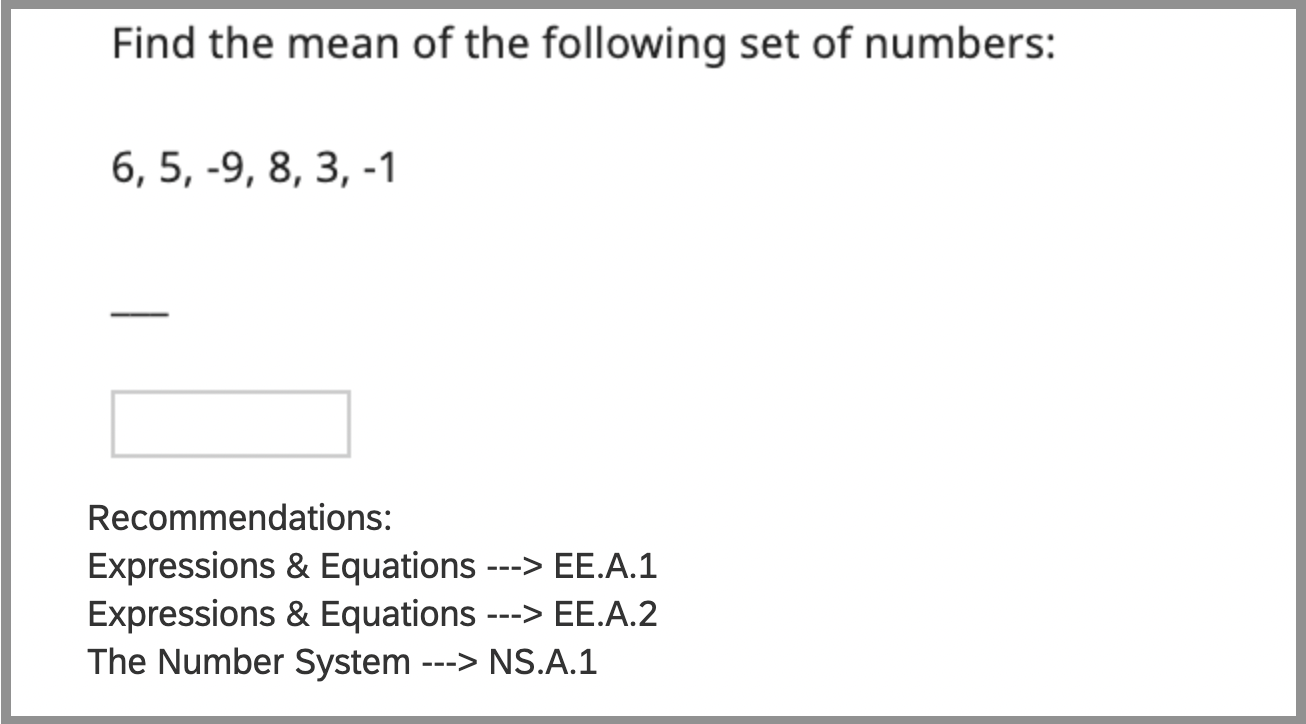}
    \caption{An example of used problems}
    \label{fig:example_questions_2}
    \vspace{0mm}
\end{figure}

In order to compare the performance with and without AI involved, we designed two versions of a survey. The first version followed the design described above, while the second version added a message recommending three skills, generated by the machine learning model and ordered by similarity score, which were shown under the math question and before the five category skill selection interface as shown in Figure \ref{fig:example_questions_2}.
Before the online experiment, we obtained IRB approval. During recruitment, we sent out a registration form to undergraduate students at a public university in the United States. We were able to recruit 22 students for the study. The experimental session took place over a 50-minute synchronous Zoom session due to the pandemic, including a 3-minute introduction in the main room, a 2-minute break in the demonstration room for the different versions of the survey, and 45 minutes for skill tagging. Some students may not have been able to finish all 30 questions within the allotted time. The students were randomly split into two equally-sized groups of 11 for the two surveys: one with AI recommendations(experimental group) and the other without(control group). To reduce the incentive for students to game the survey, completing it without authentic engagement, we required them to return to the main Zoom room and stay there until the end of the experiment.   

\subsubsection{Experiment Analysis}
We will use accuracy, recall@n, and precision@n as evaluation metrics. For a response to be considered accurate, the skill(s) selected by the study participant must exactly match the true skills. Since people do not necessarily select the same number of skills for each question, we will calculate the recall@n and precision@n, where $n$ refers to the number of skills selected and varies across different candidates and questions. We will also present the recall@n and precision@n performance of the algorithm alone, depicting how the tagging would be performed without human collaboration. 

We will aggregate the performance metrics at three levels: averaged per question, per person, and per response. We have a list of responses, each corresponding to a pair of person and question. For example, to obtain the metrics at the per question level, we will first average the metrics for a given question across participants, and then average all the individual question metrics together. 

We will report the results for the experimental and control groups and compare them using appropriate statistical tests.

Each skill has several levels based on common core skills. For example, EE.A.1 corresponds to the skill ``Expressions \& Equations -$>$ Apply and extend previous understandings of arithmetic to algebraic expressions -$>$ Write and evaluate numerical expressions involving whole-number exponents." To compare how much participants follow the recommendations, we will test whether their choices overlap with the recommendations at different levels. For instance, if the three recommendations are NS.C.7, NS.C.6, and SP.A.3, then if a participant chooses either NS or SP, we count that as 1 at the first level. We replicate this step for the second and third levels as well. Since we have 11 participants for each group, we can calculate the overlap rate for each question by counting the percentage of participants who choose the same answers as the ones recommended by the AI (e.g., a 50\% overlap rate means that half of the participants choose the same answers as the AI recommended). After we have obtained the overlap rates for each question at different levels, we will use an independent t-test to compare whether there is a significant difference between the experimental and control groups. We also conduct similar calculations by each person.

In addition to collecting responses, the survey also collects some metadata, such as the click count for each response per person.

\section{Results}
After calculating the time and recall/accuracy, the results show that the experimental group had a lower accuracy (0.115) than the control group (0.176) at all three levels, with the difference being statistically significant (p-value < 0.05) at the per question and per response levels. The experimental group also had lower recall@n and precision@n than the control group at all three levels, but only the difference in precision@n at the per-response level was significant. In terms of the time it took to tag the skills, the experimental group took significantly less time (23.5 seconds per problem) than the control group (44 seconds per problem). There was not much difference between the three levels, but the algorithm's recall@n and precision@n were worse than those of both the control and experimental groups.

\begin{table}
% \vspace{-3mm}
\caption{Online experiments Results}
\label{tab:item2skill_problem_cls}       % Give a unique label
\begin{tabular}{lccccccc}
\hline\noalign{\smallskip}

condition & metric & experimental & control & experimental size & control size & p-value \\
\noalign{\smallskip}\hline\noalign{\smallskip}
\multirow{5}{*}{\centering{per question}} & accuracy &0.115 &0.176 &30 &30 & 0.014\\
& time (s) &23.557 &44.003 &30 &30 &1.178$\times10^{-5}$\\
& acc / time ($10^{-3}/s$) &8.867 & 10.383&30 &30 &0.236\\
& recall@n &  0.38& 0.426&30 &30 &0.159\\
& precision@n & 0.273 & 0.318 &30 &30 & 0.081\\
\hline\noalign{\smallskip}
\multirow{5}{*}{\centering{per person}} & accuracy &0.115 &0.176 &11 &11 & 0.117 \\
& time (s) & 707& 1309& 11&11 &2.2$\times10^{-4}$ \\
& acc / time ($10^{-4}/s$) & 1.746&1.367 & 11&11 & 0.248\\
& recall@n &0.386  & 0.418&11 &11 &0.267\\
& precision@n &0.266  & 0.294&11 &11 &0.224\\
\hline\noalign{\smallskip}
\multirow{5}{*}{\centering{per response}} & accuracy &0.115 &0.176 &330 &330 &0.004 \\
& time (s) &23.557 &44.711 &330 &323 & 1.943$\times10^{-11}$\\
& acc / time ($10^{-3}/s$) &14.033 &17.320 &330 &330 & 0.182\\
& recall@n & 0.38 & 0.426&330 &330 &0.067\\
& precision@n & 0.292 & 0.336 &330 &330 & 0.046\\
\hline\noalign{\smallskip}
% & recall@1 & 0.225 & & & & &\\
% & recall@3 & 0.45 & & & & &\\
\multirow{2}{*}{\centering{algorithm (per question)}} & recall@n & 0.335 & & & & &\\
& precision@n & 0.236 & & & & &\\

\noalign{\smallskip}\hline
\end{tabular}
%    \begin{tablenotes}
% The p-values are calculated by independent t-test, except for the accuracy of per response(chi-square) and time of per response(independent t-test).
%    \end{tablenotes} 
% \vspace{-5mm}
\end{table}

\begin{table}
\vspace{-3mm}
\caption{Mean overlap rate with AI recommendations with p-value of independent t-test between experimental and control groups by \textbf{question} at different skill levels}
\label{tab:meanoverlap}       % Give a unique label
\begin{tabular}{lcccc}
\hline\noalign{\smallskip}
\multirow{2}{*}{Level} & Mean Overlap Rate & Mean Overlap Rate & \multirow{2}{*}{p-value} \\
& Experimental & Control & \\
\noalign{\smallskip}\hline\noalign{\smallskip}

Level 1 (e.g., NS) & 0.95 (SD=0.11) & 0.84 (SD=0.22) & 0.006 \\
Level 2 (e.g., NS.C) & 0.88 (SD=0.14) & 0.65 (SD=0.28) & $<$0.001 \\
Level 3 (e.g., NS.C.7) & 0.81 (SD=0.16) & 0.41 (SD=0.30) & $<$0.001 \\
\noalign{\smallskip}\hline
\end{tabular}
%\vspace{-6mm}
\end{table}

\begin{figure}
    \vspace{-5mm}
    \centering
    \includegraphics[width=0.9\linewidth]{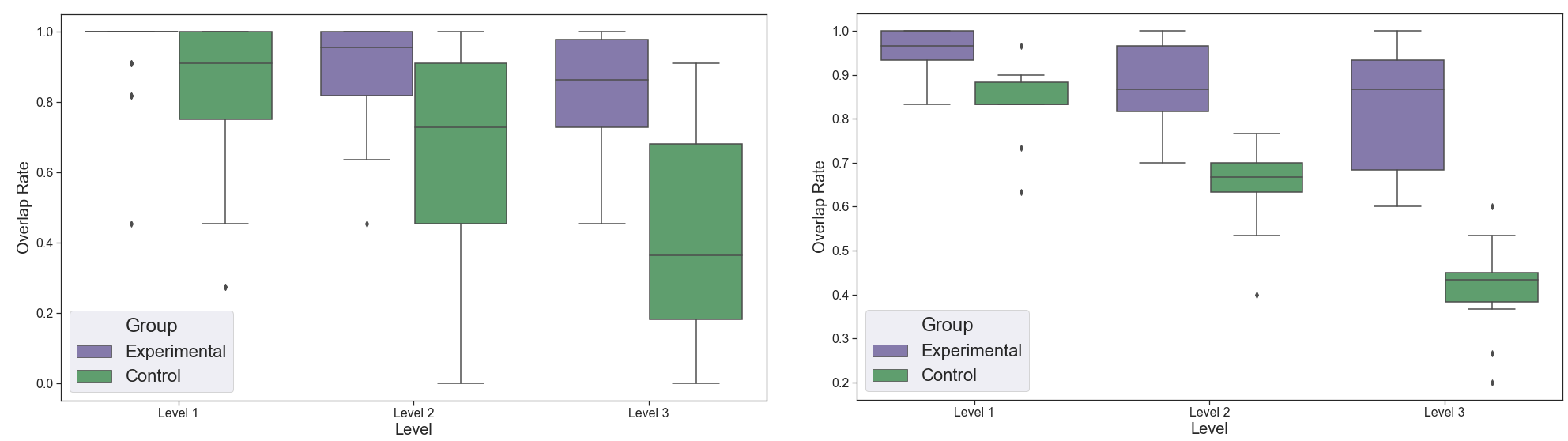}
    \caption{Boxplots of Overlap Rate by Questions (left) and by Person (right) at Different Levels }
    \label{fig:boxplotoverlap}
    \vspace{-6.5mm}
\end{figure}

Next, to explore how people in the experimental group interacted with the recommendations, we calculated the overlap rate between human choices and recommendations in the two groups. Overall, the overlap rates between the experimental and control groups were all statistically different at a significance level of 95\%, meaning that, as expected, participants in the AI condition exhibited more overlap with the AI recommendations than the control condition. Meanwhile, as the level increased, the overlap rate in both groups decreased, but the difference in the mean overlap rate between the two groups increased (see Table ~\ref{tab:meanoverlap}). 
This change indicates that there were more disagreements between humans and AI at finer granularity levels in this case. According to Figure ~\ref{fig:boxplotoverlap} (left), at a coarser level (e.g., Level 1), the range in both groups was smaller when compared to the other two finer levels. Within each level, the experimental group's range was smaller than the control group's, which shows that there was more consistency between the experimental group and the recommendations. This analysis reveals that participants with recommendations (experimental group) were more likely to follow them. When looking at the personal level, Figure ~\ref{fig:boxplotoverlap} (right) indicates that the gap between the experimental and control groups increased as the tagging granularity level of analysis became finer. The 25th percentile to 75th percentile range did not even overlap, which shows the significant influence of recommendations.
% \begin{figure}
%     \centering
%     \begin{subfigure}{0.45\textwidth}
%         \includegraphics[width=\linewidth]{boxplot_rate.png}
%         \caption{Boxplot of Overlap Rate by Questions at Different Levels}
%         \label{fig:boxplotoverlap}
%     \end{subfigure}
%     \hfill
%     \begin{subfigure}{0.45\textwidth}
%         \includegraphics[width=\linewidth]{boxplot_rate_person.png}
%         \caption{Boxplot of Overlap Rate by Questions at Different Levels}
%         \label{fig:boxplotoverlapperson}
%     \end{subfigure}
% \end{figure}

% \begin{figure}
%     \centering
%     \includegraphics[width=0.45\linewidth]{boxplot_rate_person.png}
%     \caption{Boxplot of Overlap Rate by Person at Different Levels }
%     \label{fig:boxplotoverlapperson}
%     % \vspace{-6mm}
% \end{figure}

To understand the behavior of participants while answering questions, we also analyzed the click count for each question. Since each question had multiple inputs (people), we chose the median for each question. The mean of the median click count for each question in the experimental group was 4.4, while in the control group, it was 5.7. In our design, people had to click at least twice to move on to the next question. 26.7\% of the clicks in the experimental group and 15.4\% of the clicks in the control group were less than three clicks.

\section{Discussion}
Our work advances the understanding of AI and human interaction, particularly when the AI's capacity is not as good as humans, which is typical in most AI applications at this juncture. When comparing the AI+Human and Human-only groups, the biggest benefit is time-saving. The AI+Human group saved almost half the time for the same tasks. After exploring the choice patterns in both the experimental and control groups, we believe the AI recommendations impact the speed-up effects in the experimental group. The overlap rate between the recommendations and human in the experimental group is statistically significantly higher than in the control group at every level, meaning that the recommendations were persuasive, at least to the participants in our design.  

In terms of accuracy, the AI+Human group performed better than the prediction from the AI, but still worse than the Human-only group. Humans defer more to the AI and reduce independent thoughts, which results in less time, but they are still able to detect outliers that are very far from Human expectations. This explains why the Human+AI group performed better than the AI-alone predictions and worse than the Human-only group. For example, for one question ``Find the mean of the following set of numbers: 6, 5, -9, 8, 3, -1", the overlap rate between the AI recommendation and human choice was 0.45, compared to the median rate of 1 at level 1 by question. The recommendation indicated that the skills fell under the EE (Expressions \& Equations) and NS (Number System) categories, but the correct category was SP (Statistics \& Probability), so the participants decided to reject all the recommendations and choose the answers they believed to be correct.

According to this specific case, we observed an average effect. Specifically, the performance of the AI+Human group was between that of the AI prediction and the Human-only group in terms of both accuracy and time cost. This raises an interesting question not only in AI in education but also more broadly in the workforce with both AI and humans. If the performance of the AI+Human group or the AI prediction is worse than that of the Human-only group, what will the trade-off look like? Will decision-makers choose to take the speed-up or maintain higher accuracy? It is difficult to discuss this without specific contexts, but we do see cases in different fields such as healthcare trying to speed up operations with AI \cite{abdullah2020health}. 

If the AI's performance more closely rivaled humans', we might expect an ensemble effect whereby the combination of the two leads to a superior accuracy rather than an accuracy that is the average of the two, which we observed in our study. In that case, we can speculate that human taggers could continue to correct obvious mistakes made by AI due to lacking causal inference and other human faculties, and that there would be fewer less obvious, uncaught errors. Desmond et al. \cite{desmond2021increasing} noted that their AI system with humans achieved better results in both speed and accuracy(0.79) than the Human-only group (0.72) although with low accuracy from AI-only (0.44). Jarrahi \cite{jarrahi2018artificial} argues that the future of AI and humans is for intelligence augmentation, which emphasizes that each side can bring its own strengths to decision-making processes.

As AI tools, like ChatGPT, gain increased popularity and accessibility, there's a growing interest in comparing AI and human efficacy in educational tasks and exploring  avenues for AI-Teacher and AI-Student collaboration. However, nascent evaluations of ChatGPT in educational settings are finding subpar performance of the model in comparison to human experts. For instance, Pardos and Bhandari \cite{pardos2023learning} leveraged ChatGPT to generate hints for Algebra courses, with 30\% of the hints produced failing manual quality checks. Similarly, Wang and Demszky \cite{wang2023chatgpt} engaged math teachers to assess the zero-shot performance of ChatGPT in tasks like identifying highlights for good instructional strategies within math classroom transcripts. However, 82\% of the model’s suggestions repeated the teachers' suggestions. Given the ongoing integration of ChatGPT by some teachers, it's crucial to provide suggestions or even evidence on where ChatGPT can effectively enhance teaching and learning experiences, clarifying how AI can best support educational outcomes. These findings, in combination with our results, suggest that premature collaboration with AI may result in degraded quality of outcome and potentially lower educational quality.

\section{Limitations and Future Work}
First, during the tagging process, students were not trained in the skill tagging tasks. Skill tagging is typically not performed by students, but rather by domain experts. Although we asked college students in STEM during the recruitment, they don't know the rules followed by CK12 taggers. We observed similar designs in other studies where the participants were usually from Amazon Mechanical Turk \cite{desmond2021increasing,lai2019human}. However, compared to those tasks, skill tagging might be more cognitively different. Thus, future research could ask experts to participate in these tasks and observe how the results and interactions will change. 

Second, we only have a small sample(N=22) and do not know the participants' level of math knowledge. There could be a selection bias since we sent out our recruitment on several STEM classes at one public university. Meanwhile, as mentioned above, people's confidence will impact the interaction between humans and machines. The absence of participants' knowledge levels also limited us from exploring and explaining their choice, though since they are undergraduates we can expect an above K-12 level of math knowledge. 

Third, questions may have more than one skill associated with them, but our algorithms always provided three recommendations without confidence probability. For some questions, the recommendations may not have a strong relationship but may still be listed as the third recommendation. Thus, in the future, the design could be expanded in other aspects, such as providing recommendations based on different accuracy levels or providing extra information, such as confidence levels. This may help us understand in what kinds of circumstances will humans challenge the recommendations from AI.

Fourth, we discussed that although humans can still override bad recommendations, we did not observe ensemble effects. Some papers have explored the possibilities of ensemble effects. Thus, it would be interesting to investigate what kinds of factors can activate ensemble effects and maximize efficiency as well as accuracy with AI.

\section{Conclusion}
This study presents an experiment exploring AI+human collaboration and differences between humans with and without AI assistance in skill tagging tasks. The results show that taggers with AI assistance save almost 50\% time (p $<$$<$ 0.01) compared to without AI, but sacrifice 7.7\% recall (p = 0.267) and 35\% accuracy (p= 0.1170). We observed average effects, where AI+human's achievements are between humans only and AI only, particularly in terms of time and accuracy. We also observed that participants in the AI condition were highly influenced by the recommendations but did not follow them blindly. The recommendations led to skill choices that were significantly different from the Human-only condition, even at the most coarse grain skill level of analysis.

At present, the educational field is rapidly embracing AI, particularly in this new age of Large Language Models like ChatGPT. While there are examples of industries adopting Human-AI collaboration approaches in the service of efficiencies and at the expense of quality, this is not an appealing trade-off given the values of public educational institutions. Human-AI collaboration may be a feasible strategy based on financial considerations and workload reduction, particularly in tasks that would not be possible to scale without AI; however, it is still common at this stage for AI to be less accurate than humans, particularly in conceptually sophisticated tasks. When AI is not yet at a sufficient level of accuracy relative to the human expert, our finding suggests that collaboration can result in a lower quality outcome than the human expert alone. Researchers should therefore continue to rigorously measure and monitor at what point and for which tasks AI collaborations produce a net educational benefit before they are widely adopted. 

%to this subject and for future implementation, highlights some promising aspects of AI, but more importantly, provides insights into the current benefits and downsides of integrating AI in an educational task.

% \section{SIGCHI Extended Abstracts}

% The ``\verb|sigchi-a|'' template style (available only in \LaTeX\ and
% not in Word) produces a landscape-orientation formatted article, with
% a wide left margin. Three environments are available for use with the
% ``\verb|sigchi-a|'' template style, and produce formatted output in
% the margin:
% \begin{itemize}
% \item {\verb|sidebar|}:  Place formatted text in the margin.
% \item {\verb|marginfigure|}: Place a figure in the margin.
% \item {\verb|margintable|}: Place a table in the margin.
% \end{itemize}

% %%
% %% The acknowledgments section is defined using the "acks" environment
% %% (and NOT an unnumbered section). This ensures the proper
% %% identification of the section in the article metadata, and the
% %% consistent spelling of the heading.
% \begin{acks}
% \end{acks}

\bibliographystyle{ACM-Reference-Format}
\bibliography{main}

% %%
% %% If your work has an appendix, this is the place to put it.
% \appendix









\end{document}